\documentclass[twocolumn]{aastex61}

\received{---}
\revised{---}
\accepted{---}
\submitjournal{ApJ}

\shorttitle{Hot plasma in a quiescent active region}
\shortauthors{Ishikawa et al.}

\begin{document}

\title{Evaluation of hot plasma in a quiescent solar active region with \textit{RHESSI}}
\title{Hot plasma in a quiescent solar active region as measured by \textit{RHESSI}, \textit{XRT}, and \textit{AIA}}

\correspondingauthor{Shin-nosuke Ishikawa}
\email{s.ishikawa@isee.nagoya-u.ac.jp}

\author{Shin-nosuke Ishikawa}
\affil{Institute for Space-Earth Environmental Research, Nagoya University, Nagoya, Aichi 464-8601, Japan}

\author{S\"am Krucker}
\affil{Institute for Space-Earth Environmental Research, Nagoya University, Nagoya, Aichi 464-8601, Japan}
\affil{Space Sciences Laboratory, University of California, Berkeley, CA 94720-7450, USA}
\affil{Institute for Data Science, School of Engineering, University of Applied Sciences and Arts Northwestern Switzerland, 5210 Windisch, Switzerland}



\begin{abstract}
This paper investigates a quiescent (non-flaring) active region observed on July 13, 2010 in EUV, SXR, and HXRs to search for a hot component that is speculated to be a key signature of coronal heating. We use a combination of RHESSI imaging and long-duration time integration (up to 40 min) to detect the active regions in the 3-8 keV range during apparently non-flaring times. The RHESSI imaging reveals a hot component that originates from the entire active region, as speculated for a nanoflare scenario where the entire active region is filled with a large number of unresolved small energy releases.  
An isothermal fit to the RHESSI data gives temperatures around $\sim$7 MK with emission measure of several times 10$^{46}$ cm$^{-3}$. 
Adding EUV and SXR observations taken by AIA and XRT, respectively, we derive a differential emission measure (DEM) that shows a peak between 2 and 3 MK with a steeply decreasing high-temperture tail, similar to what has been previously reported. 
The derived DEM reveals that a wide range of temperatures contributes to the RHESSI flux (e.g. 40\% of the 4 keV emission being produced by plasma below 5 MK, while emission at 7 keV is almost exclusively from plasmas above 5 MK) indicating that the RHESSI spectrum should not be fitted with an isothermal.
The hot component has a rather small emission measure ($\sim$0.1\% of the total EM is above 5 MK), and the derived thermal energy content is of the order of 10\% for a filling factor of unity, or potentially below 1\% for smaller filling factors. 
\end{abstract}




\section{Introduction}

To investigate energetics in the solar corona and explore how the corona maintains its high temperature, it is important to detect and quantitatively evaluate the hot ($>$5~MK) plasma component above the typical coronal temperature (2--3~MK) in active regions \citep[e.g.,][]{klimchuk2009}.
Although solar flares heat the coronal plasma well above 10~MK, it is known that the energy released by flares individually is not significant in the coronal energy balance \citep[e.g.,][]{shimizu1995}. 
Therefore, the important next steps are studies of the hot component during quiescent times when no individually resolved flare is detected. Simulations suggest that a quasi quiescent hot plasma component could be produced by a superposition of the large number of discrete, but temporally overlapping, compact impulsive energy releases that are distributed over the entire Sun \citep{cargill2004,klimchuk2008,bradshaw2011}, called nanoflares, originally introduced by \cite{parker1988}.

From the observational side, temperature structures of active region plasmas have been investigated using observations at various wavelenghts, such as extreme ultraviolets \citep[EUVs, i.e.,][]{warren2012,brosius2014,parenti2017} and 
soft X-rays \citep[SXR, i.e.,][]{parkinson1975, peres2000, orlando2001, orlando2004, reale2009a,schmelz2009a,delzanna2014}. 
In these studies the differential emission measures (DEMs), temperature derivatives of emission measures, are reconstructed from the observations. Deriving the DEM from this limited set of observations is not providing a unique solution, and calibration uncertainties can therefore significantly influence the result. The temperature range over which the DEM is reconstructed should be carefully selected.
Hot plasmas above  $>$5~MK emit several EUV lines \citep{young2007} and SXR emission \citep{golub2007} and therefore EUV and SXR observations in principle provide good diagnostics of hot plasmas. 
However, it has been pointed out that a combination of EUV and SXR observations is not enough to evaluate the hot plasma with current instruments such as EUV Imaging Spectrometer \citep[EIS,][]{culhane2007} and X-ray Telescope \citep[XRT,][]{golub2007} onboard the Hinode satellite, except at times when large flares occurred \citep{winebarger2012}. During non-flaring times, EUV lines sensitive to the hot plasma are simply too faint to be detectable by EIS, and the temperature response of XRT is too wide and therefore dominated by the much larger emission measures of cooler plasma. 

To properly detect the high-temperature tail, hard X-rays (HXR, X-ray emissions with a few keV and above hereafter) are an essential diagnostic tool. HXRs are produced by bremsstrahlung, but only the tail of the electron distribution has  enough energy to make photons in the HXR range. Hence, HXR bremsstrahlung observations are biased towards the hottest plasma making them the ideal diagnostic of the hottest temperatures. \cite{schmelz2009b} showed that the hot component is highly constrained by an HXR upper limit derived from an instrument background of the \textit{Reuven Ramaty High Energy Solar Spectroscopic Imager} \citep[\textit{RHESSI},][]{lin2002} satellite. Therefore, a combination of HXR with other wavelength bands is essential to accurately evaluate a DEM with a wide temperature range that includes the hot component. \cite{mctiernan2009} analyzed RHESSI observations of day-night transitions during non-flaring times, and found HXR emissions from 5--10~MK plasma demonstrating that a hot component is typically present, at least within the sensitivity range of RHESSI. Their work was a spectroscopic analysis using the day-night transition for an accurate non-solar background subtraction, but no imaging was performed. The detected emissions, however, are thought to come from active regions. 
\citet{reale2009b} compared a RHESSI observations with SXR observations by Hinode/XRT, and found that RHESSI mainly detects emission from a hot plasma around 6--8~MK and XRT could mainly observe cooler plasmas around 2--2.5~MK, indicating the existence of a hot component as envisioned by nanoflare coronal heating models.  

More recently, advances have been made with high sensitivity HXR focusing optics telescopes. HXR focusing optics has much improved sensitivity compared to RHESSI's indirect imaging method by achieving large effective areas and a low non-solar background. The Focusing Optics Solar X-ray Imager (FOXSI) sounding rocket experiment applied this technique for the first time to the solar observation \citep{krucker2014}. The first flight had a moderate sensitivity that was not high enough to detect HXR emissions from a non-flaring active region, but the observations nevertheless strongly constrained the hot component in active regions \citep{ishikawa2014}. During the second launch of the FOXSI sounding rocket which provided a much improved sensitivity, HXR emissions from a quiescent active region have been detected \citep{ishikawa2017}, corroborating the existence of the hot component even above 10~MK. The NuSTAR satellite \citep{harrison2013} has provided further hard X-ray focusing observations giving new insights into the existence of a hot component. As NuSTAR has not been designed for solar observations, the observations so far are limited by short effective exposure times, and only upper limits of the quiescent component above 5 MK have been derived so far \citep{hannah2016,grefenstette2016}. The availability of all these new hard X-ray focusing observations have triggered several simulation studies, and initial results of nanoflare models are constructed to explain those observations \citep{barnes2016a,barnes2016b,marsh2018}. 

In this paper we revisit the RHESSI data taking advantage of the possibility to image and integrate in time by almost an hour to compensate for RHESSI's moderate effective area. We report on hard X-ray emissions in the range from 3 to 8 keV from an active region on July 13, 2010 during time intervals without any individual X-ray flares. RHESSI was able to successfully obtain HXR images of this quiescent active region, and by combining with EUV and SXR observations, it was possible to measure the differential emission measure distribution and compare the core component with the hot tail of the distribution. 

\section{Observations}

\begin{figure*}[ht!]
\plotone{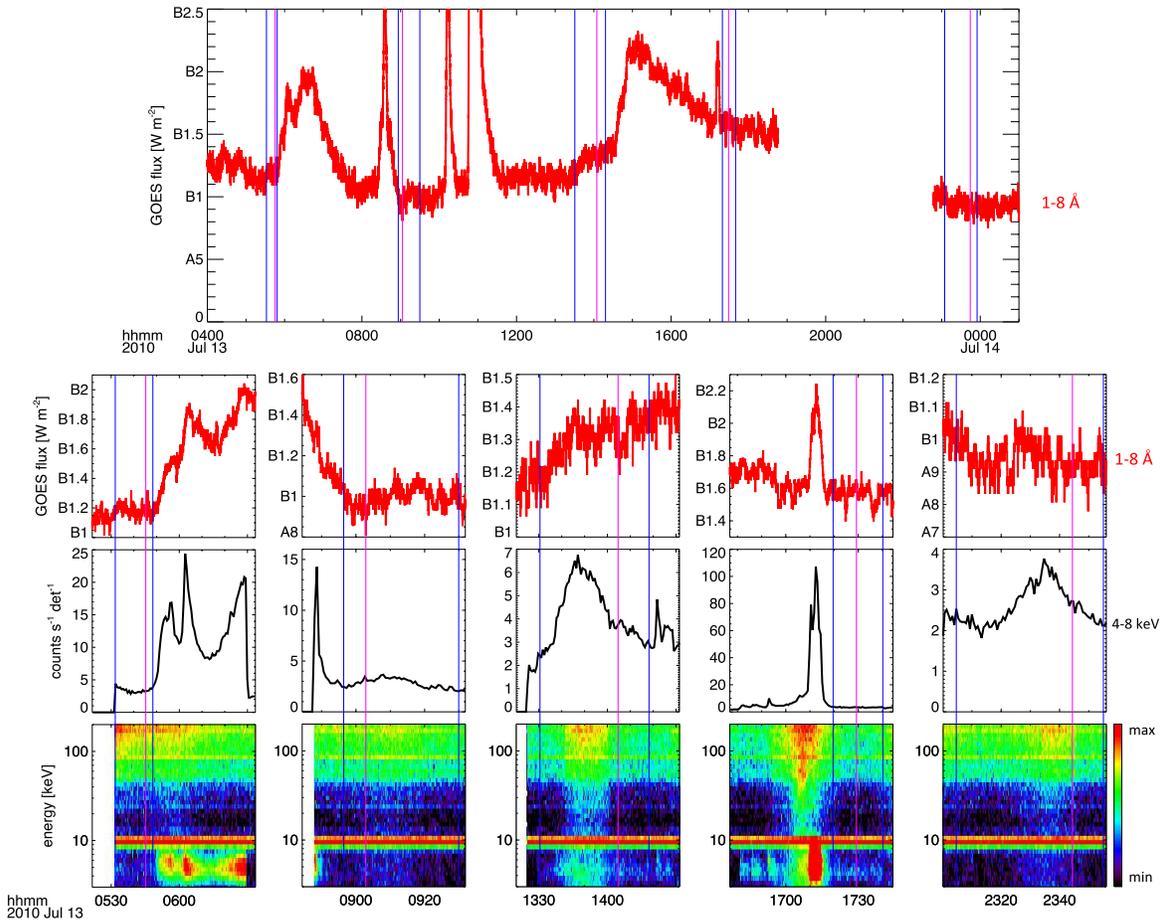}
\caption{Overview plot of \textit{GOES} SXR and RHESSI HXR lightcurves on July 13, 2010.  Top: GOES 1--8~\AA {} lightcurve in linear scale.  Bottom: GOES 1--8~\AA {} lightcurve, RHESSI 4--8 keV lightcurve, and RHESSI spectrogram for the 5 selected orbits (see text for the details). The data used for the spectrogram plots are the RHESSI count rates and the same logarithmic  scaling is used for all plots with the minimum and maximum corresponding to 0.3 and 5 counts per second per detector. Non-solar background emissions change during RHESSI's orbit and can be best identified by their high energy signal around 100 keV in the spectrogram plot. Blue lines show the time intervals for the DEM analysis, and magenta lines give the time of the XRT observations with the Al-thick filter.  \label{fig:lightcurves}}
\end{figure*}

Our study presented here does not intend to be a statistical search, but we wanted to find an apparently quiescent, non-flaring active region that is detected close to the RHESSI sensitivity limit. To be able to compare the RHESSI results with extreme UV and soft X-ray observations, we restricted our search for periods after February 2010 when observations of the \textit{Atmospheric Imaging Assembly} onboard the \textit{Solar Dynamic Observatory} satellite \cite[\textit{SDO/AIA},][]{lemen2012} started. In addition, only times when Hinode/XRT was running in a multiple filter configurations especially designed to observe active region were considered. To have good calibration of the RHESSI data and minimal effects of radiation damage, we further restricted the search to times within a few months after the second RHESSI anneal that was completed by the end of April 2010. To facilitate RHESSI imaging, we looked further at times when a single active region was dominating the total X-ray flux. This simplified RHESSI imaging as a single source is much easier to image for RHESSI than two widely separated sources from different active regions. A good candidate for our study was found to be 
NOAA AR 11087, a single active region seen on the disk on July 13, 2010 that has good coverage by RHESSI, AIA and XRT. 
The upper panel of Fig.~\ref{fig:lightcurves} shows the \textit{GOES} X-ray lightcurve in linear scale for part of the selected day. The solar activity was low with the X-ray background flux being around B1 level. Several A- and B-class microflares occurred throughout the day, but there are also times in between microflares without obvious activity in the GOES light curve. 
The RHESSI spacecraft has a day-night cycle providing uninterrupted solar observations of up to 60 minutes per 96 minute orbit. Avoiding times when RHESSI crossed the South Atlantic Anomaly, we selected 5 RHESSI orbits when XRT ran a standard active region observation program with two filter configurations, the Al-mesh and Ti-poly filters, and an image with the Al-thick filter was taken every hour. GOES and RHESSI data for the 5 selected orbits are shown in the bottom panel of Fig.~\ref{fig:lightcurves}. The RHESSI data is shown as a lightcurve (4-8 keV), but also as spectrogram plots covering the energy range from 3 to 200 keV. In the spectrogram plot where the observed HXR count rate is shown in color in a 2D representation as a function of time and energy, the RHESSI non-solar background variations can be clearly identified by following the high-energy time variations. The observed variations reflect the spacecraft latitude with the background increasing towards higher latitudes. We selected intervals for RHESSI imaging spectroscopy analysis (blue lines) that exclude times of obvious X-ray microflares seen by GOES and RHESSI. The emissions in the selected time intervals are dominated by non-solar background without an obvious solar signal. In the following we describe our efforts to use long-time integration RHESSI imaging to search for a hidden solar signal within these intervals.

The selected time intervals for RHESSI imaging have a duration between 17 and 55 minutes, much longer than typically used when imaging flare emissions where duration below 1 minute are generally used. However, similar duration integrations have been used for RHESSI imaging in the gamma-ray range \citep[e.g.,][]{hurford2006}. The long time integration allows us to increase statistics for these times of very weak emissions. We reconstructed RHESSI images for those intervals with the standard CLEAN algorithm, as it appears to be the most robust algorithm for low-statistics imaging. Despite the long integration, counting statistics are still low with typically a few thousand counts per detector per interval between 3.5--6.5~keV, of which at least half are non-solar background counts (i.e. unmodulated counts). These are low numbers compared to RHESSI images with excellent statistics ($\sim$10$^5$ counts/detector), but similar to statistics available for the largest gamma-ray flare \citep{hurford2006}, although at much higher background. 

\begin{figure*}[ht!]
\plotone{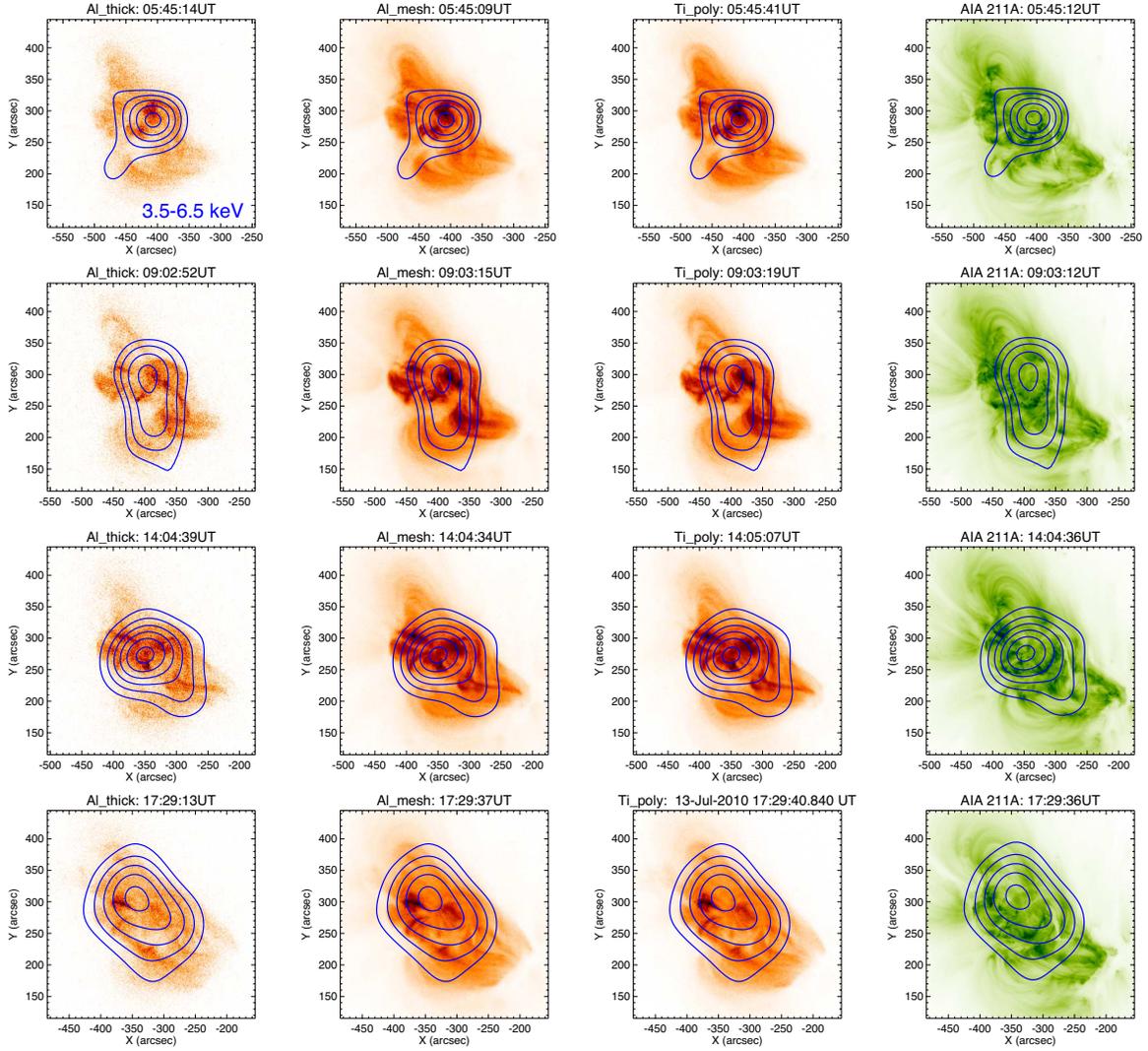}
\caption{RHESSI 3.5--6.5 keV images (blue contours at 61$''$ FWHM) overlaying the XRT images.  The columns correspond to the time intervals for the orbits 1--4.  
The rows correspond to the XRT filter configurations, with the Al-thick, Al-mesh,  Ti-poly filters, and AIA 211\AA.\label{fig:images}}
\end{figure*}

For all but  interval 5, RHESSI 3.5--6.5~keV images show an extended source covering the entire active region (interval 3 and 4), or at least a large part of the active region (interval 1 and 2). For interval 5, no significant modulation was seen in the RHESSI data, indicating that for this interval the solar signal relative to the background was too weak for imaging, or even absent. RHESSI images for the other intervals are shown as blue contours in Fig.~\ref{fig:images} overlayed on the XRT images. Each column corresponds to one of the selected time intervals, and rows correspond to the different XRT filters. As the source is extended, only the coarse subcollimators 6 to 9 with natural weighting (equal weighting resulting in an angular resolution of 61$''$ FWHM) are used in the reconstruction of the images shown in Fig.~\ref{fig:images}.

\begin{figure*}[ht!]
\plotone{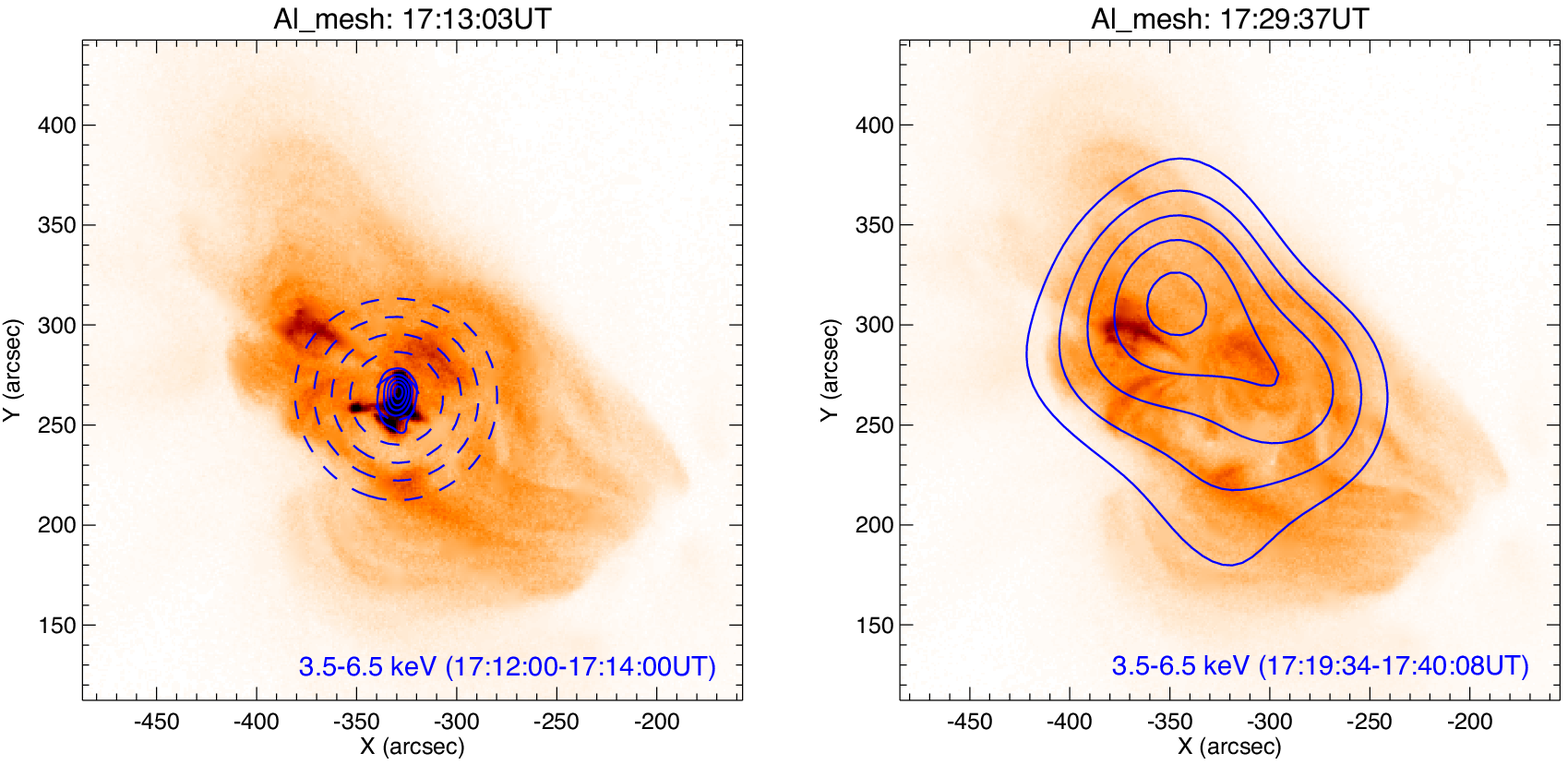}
\caption{RHESSI images of the active region overlaid on the XRT image during an occurrence of a microflare and at the quiescent time interval for the DEM analysis.  Left: With microflare.  Blue solid contours show the image obtained with subcollimators 3--9, and dashed contours show the image with only coarse grids same as used in Fig.~\ref{fig:images} and the DEM analysis (subcollimators 6--9).  Right: Orbit 4 time interval after the microflare.  \label{fig:compareimage}}
\end{figure*}

To further corroborate the extended nature of the quiescence source, we compared the imaging result of interval 4 to the microflare that occurred a few minutes earlier (see Fig.~\ref{fig:compareimage}). The microflare comes from a compact source located in the southern part of the quiescent source and it is co-spatial with an XRT brightening. In addition to the high-resolution image of the microflare, the dashed contours show the microflare image reconstructed with the same parameters as we use for imaging the quiescent active region. This clearly confirms that the quiescent source is not related to the microflare occurring early, but it is an independent component from the entire active region. 

In a second step, we used RHESSI imaging to determine the X-ray flux of the quiescent active region at 1 keV bin size.  As the non-solar background signal is not modulated by the RHESSI imaging system, using imaging to determine the flux has the advantage that the non-solar background is not contributing to the reconstructed image. Hence, imaging provides a simple background subtraction without selecting a pre-flare time interval to determine the background. We obtain RHESSI images at 1 keV bin size using subcollimator 9 and then calculate the fluxes by summing all pixels within the 50\% contours and multiplying by a factor of 2 to account for the wing of the clean beam (Table~\ref{table:table}). Before we present a differential emission measure analysis in the next section, we briefly mention here the standard approach of fitting the RHESSI fluxes with a single-temperature model. Using only the lowest three energy bins (3.5--6.5~keV) gives values between 5.3--7.1~MK with EM between  $\sim$3 to 13$\times 10 ^{45}$~cm$^{-3}$, while including the 6.5 to 7.5 keV bin for the three intervals with good statistics gives values slightly higher 7.1--7.4~MK at lower EM ($\sim$2 to 4$\times10 ^{45}$~cm$^{-3}$). This indicates that there is a distribution of temperature that contributes to the RHESSI observations, but a single temperature approximation is apparently  representing the data reasonably well. As we show in the following section, this will no longer be the case after we include EUV and SXR to the analysis.

\begin{table*}[ht!]
\centering
\caption{\textit{RHESSI} HXR fluxes in each \textit{RHESSI} orbit on July 13, 2010.  Isothermal temperatures derived by those \textit{RHESSI} fluxes are also shown.} \label{table:table}
\begin{tabular}{cccccccc}
\tablewidth{0pt}
\hline
\hline
Orbit & Time interval & \multicolumn{4}{c}{Flux [photons/s/cm$^{2}$/keV]}&  \multicolumn{2}{c}{Temperature [MK]}\\
 No.&  [UT]& {3.5--4.5 keV} &  {4.5--5.5 ke}V &  {5.5--6.5 keV}  &{6.5--7.5 keV}  &with 3.5--6.5~keV&with 3.5--7.5~keV\\
\hline
\decimals
1 &05:31:48--05:48:22& 16.1& 1.95& 0.222& 0.0780&6.25 &7.23 \\
2 &08:56:25--09:29:55& 9.02& 0.884& 0.0912&--&5.31 &-- \\
3 &13:30:30--14:18:10& 15.6& 2.44& 0.328&0.0814&7.06 &7.36 \\
4 &17:19:34--17:40:08& 22.0& 2.38& 0.372&0.0828&6.53 &7.02 \\
5 &23:04:40--23:55:00& $<$3.0& $<$0.1& $<$0.4&--&-- &-- \\
\hline
\end{tabular}
\end{table*}

\section{DEM analysis}
We estimated the DEM of the active region for each time interval shown in the previous section using the RHESSI, SDO/AIA and Hinode/XRT data. We use the HXR fluxes from the entire active region given in Table~\ref{table:table}, and AIA data are summed over the extent of the active region and averaged over each time interval. The XRT images are available at a lower cadence compared to the AIA images but are treated in the same way. For the Al-thick filter, a single image is available for the entire interval and we assume that it is representative of the average. We integrated the fluxes for a $360'' \times 360''$ square area to cover the whole active region and entire RHESSI HXR fluxes. AIA channels of 94, 131, 171, 193, 211 and 335~\AA {} were used to investigate a wide temperature range in the corona. For the DEM analysis, we set the temperature range of 0.6--25~MK, following to the sensitive temperature ranges of the combined observations. We processed the XRT data with an IDL procedure \verb+xrt_prep.pro+ version v2014-Jan-15 in the Solar Software (SSW) package, and used the Level-1 AIA data distributed online.  We generated temperature responses using \citet{narukage2014} calibration for XRT and calibration version V6 for AIA.
Consistency between the calibrations between XRT and AIA for our DEM analysis were checked according to \citet{boerner2014}.

For the DEM estimation, we used an IDL procedure \verb+xrt_dem_iterative2.pro+ in SSW.
This procedure assumes a spline function for the DEM, and it searches the least chi-squared solution. 
We input fixed errors of 20~\% for all the fluxes to simulate systematic errors and calculate the chi-squared value.

\begin{figure*}[ht!]
\plotone{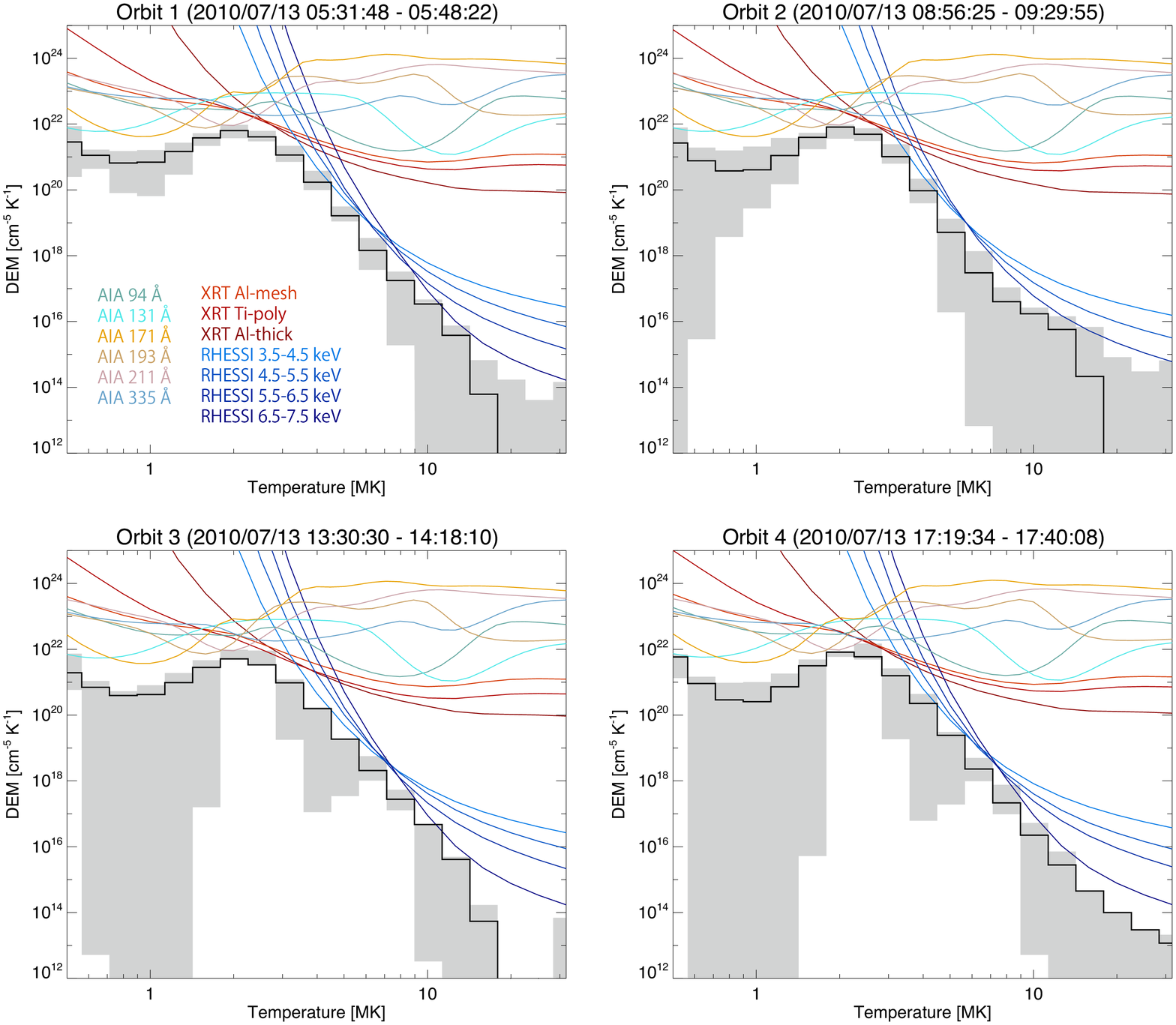}
\caption{Loci curves and DEMs for the orbits 1--4 in Table~\ref{table:table}.  The black lines show the estimated DEMs, and shaded areas correspond to the range of the solutions within 95~\% occurrence probabilities from the Monte Carlo runs. \label{fig:dem}}
\end{figure*}

\begin{figure*}[ht!]
\plotone{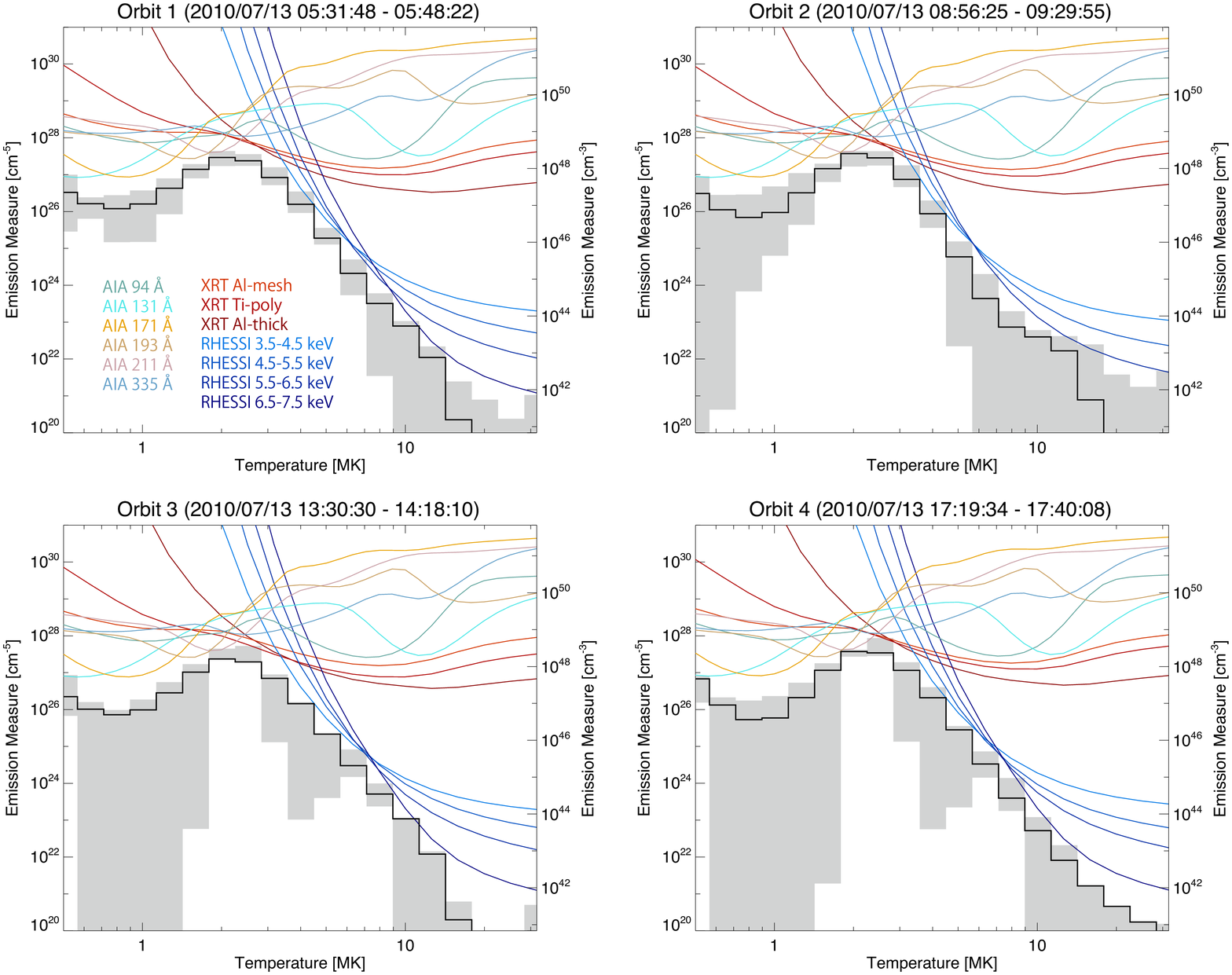}
\caption{Same as Fig.~\ref{fig:dem}, with the dimension of the emission measure.   Left ticks show emission measures per area in the unit of cm$^{-5}$ and right ticks show emission measure in the unit of cm$^{-5}$. \label{fig:em}}
\end{figure*}

To evaluate the influence of measurement errors, we calculate a set of DEM solutions from 10000 Monte Carlo runs with randomly modified input values within Gaussian distributions. The estimated DEM solutions for the time intervals in the orbits 1--4 are plotted in Fig.~\ref{fig:dem}. Plots for the emission measures per area in the cm$^{-5}$ unit are shown in Fig.~\ref{fig:em} for easier comparison with previously published results. Also, the right axis of Fig.~\ref{fig:em} show the emission measures in cm$^{-3}$. In addition to the DEM, loci curves derived from the observations and instrument responses are also shown in the figures. Each loci curve gives the emission measure of an isothermal component that is necessary to produce the observed flux.
Therefore, a valid DEM solution needs to be below all loci curves to avoid too much expected fluxes compared to the observed values. Black lines are the DEM solutions, and the shaded areas show the range of the acceptable solutions, within occurrence probability of 95~\% in the chi-squared distribution in the Monte Carlo runs. 


\begin{figure*}[ht!]
\centering
\includegraphics[width=4in]{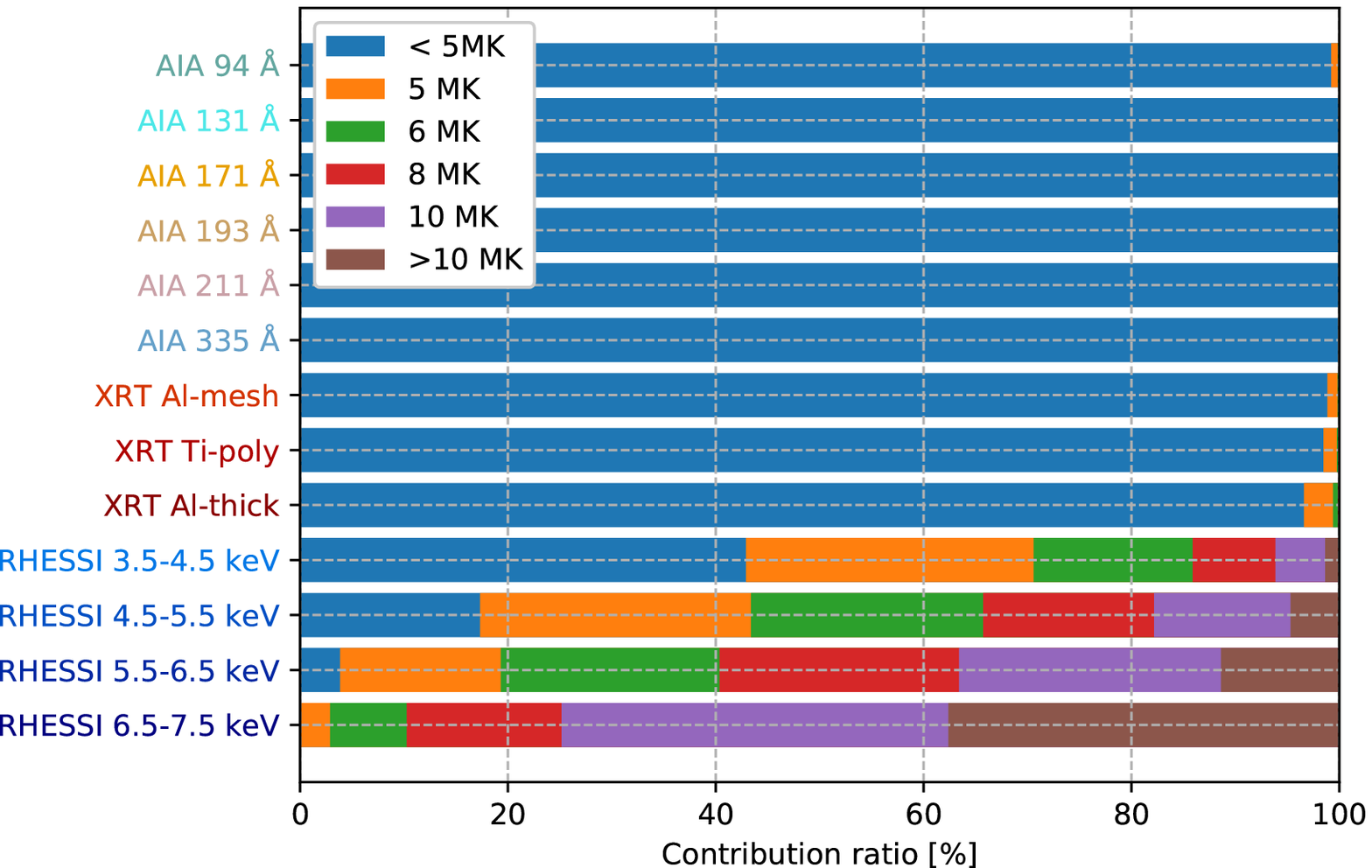}
\caption{Visual representation of our findings that the AIA and XRT telescopes detect mainly counts from the low temperature plasma, while RHESSI counts are dominated by a rather wide range of temperature of the hot tail of the distribution. The plot shows the contribution to the total flux separated by temperature for each of the different wavelength ranges for AIA, XRT, and RHESSI. Only the data for orbit 1 is shown, but the precentage are very similar for all orbits.  \label{fig:contribution}}
\end{figure*}

According to the estimated DEMs in Fig.~\ref{fig:dem}, we can clearly see that a main temperature component peaks around 2 and 2.5 MK. This main temperature component is dominating all AIA channels, and there is essentially no contribution from the hot component ($>$5~MK) to any of the AIA observations. Also the XRT observations are almost exclusively reproduced by $<4$~MK plasmas. Conversely, we cannot reproduce the HXR emissions observed by RHESSI only by the 2--2.5~MK main coronal temperature component. If we only have plasma with temperatures at 2.5~K or lower, only $<$5~\% of the HXR fluxes observed by RHESSI are expected for all the detected energy bins with the best solutions for all of the 4 time intervals. For the HXR energy range above 4.5 keV, contributions from the main temperature component are essentially absent with values below 0.3~\% for all the orbits. Fig.~\ref{fig:contribution} summarizes the contributions of the $>$5~MK hot plasma and lower temperature plasma components to each wavelength range in the case of the best DEM solution for the orbit 1 observation. The figure nicely illustrates that the contributions of the hot (5~MK and above) plasma components are dominant to the RHESSI observations, while the contributions to the EUV and SXR observations are negligible. The responses to hot emission of AIA 94 and 131~\AA {} with peaks at 8~MK and 13~MK, respectively, are too weak to significantly produce the total flux, as the loci curves are more than 2 and 3 orders of magnitude away from the shaded areas of the estimated DEMs, respectively (see Fig.~\ref{fig:em}). For the soft X-ray filters, even the XRT Al-thick observations only have a few percent contribution by the hot component. Hence, EUV and soft X-ray filter diagnostics are valuable in describing the main temperature components of active regions, but they are not suitable for the search of a hot component in quiescent active regions where X-ray imaging spectroscopy is the best diagnostic tool. 
We note that a hot plasma could be detectable by AIA if it is localized and the contribution is comparable to that of lower energies.  
For example, the contribution of the hot component is 0.78~\% for the AIA 94~\AA {} observation.  
Therefore, if this component would be localized in an area smaller than $30'' \times 30''$ (smaller than 0.7~\% area compared to the $360'' \times 360''$ region of interest), 
the hot component would be dominant in the AIA 94~\AA {} image in that area.  
However, this is not the case in this region because RHESSI imaging shows that the hot emission is extended or at least much larger than $30"$ in diameter. 

\section{Discussion}
As shown in the previous section, we simultaneously imaged a quiescent active region in HXRs, SXRs and EUVs. The combined DEM analysis reveals that emissions at these different wavelengths are not from a single isothermal component, but a range of temperatures with a hot tail are necessary to reproduce the observations. With the hot tail only observed over a rather narrow energy range, we cannot exclude that this high-energy component potentially contains a non-thermal part or could even be entirely produced by a non-thermal tail in electron distribution. The very steep nature of the observed HXR spectra however favors a thermal interpretation, which we adopt in the following discussions.

The DEM has a main component peaking between 2 and 2.5 MK with a well determined DEM with small solution ranges (small shaded areas) and emission measures between 2.3 and 8.7$\times 10 ^{48}$~cm$^{-3}$ for the 4 time intervals. Below 1~MK, our set of observations only poorly constrain the DEM. A hot component is required for all the acceptable solutions in the Monte Carlo runs up to  $\sim$6~MK for orbit 2 and $\sim$8~MK for the other orbits. These upper limits in temperature are due to the limited sensitivity of our HXR data. For even hotter plasma, RHESSI only provides upper limits of the emission measure. Ranges of the emission measures of the hot plasma above 5~MK and above for the different Monte Carlo runs shown in Fig.~\ref{fig:dem} are 
6.3--16$\times 10 ^{45}$, 
0.033--11$\times 10 ^{45}$, 
6.6--17$\times 10 ^{45}$ and 
6.3--16$\times 10 ^{45}$~cm$^{-3}$ for the orbits 1, 2, 3 and 4, respectively. 
Note that the range is largest for orbit 2 because the HXR non-detection at 7~keV resulting in large uncertainty in the quantitative estimation of the hot component.

Compared to the isothermal fit values that yielded slightly higher temperatures, these emission measures are in general higher. This is a direct implication of the isothermal assumption that ignores the contribution from lower temperature plasmas. Since the emission measure ranges of the hot component overlap for all time intervals, we found no evidence of a systematic time variation of the hot component over 18 hours. The emission measures of the hot plasma component are relatively small compared to the main component. The total emission measures derived for the four intervals by integrating over the entire temperature range from  4.8 to 8.7$\times 10 ^{48}$~cm$^{-3}$, therefore the contribution of the hot component to the total emission measure is $<$0.5~\%. 

RHESSI imaging reveals that the hot component covers almost the entire AR, suggesting a large volume of the hot component. However, RHESSI observations with its limited number of measured visibilities (Fourier components) cannot distinguish between an extended source and an apparently extended source that is composed of a large number of compact subsources. Hence, we cannot differentiate  between a scenario where hot plasma is coming from a volume-filling source or many compact small sources, such as produced by a large number of nanoflares. This makes it difficult to estimate the energy content of the hot component without making further assumptions. For a volume-filling hot component, the small emission measure indicates a much lower density of the hot component compared to the main component resulting in a total thermal energy content of the hot component relative to the main component of $<$14~\%. To get an estimate for a case with many unresolved sources, we could assume that the densities of the hot and main temperature components are similar, resulting in a volume filling factor equal to the ratio of the hot to the main emission measure component of $<$0.005, and a thermal energy content of the hot component is 1~\% or less. However, without knowing the actual filling factor, no definite statement can be made, but the hot component has in any case a much lower energy content than the main component with the value of 14~\% derived from the volume-filling assumption being an upper limit.  

In the following, we compare the findings from our joint RHESSI/AIA/XRT study to previously published results of X-ray observations of non-flaring active regions.

\subsection{RHESSI}
Compared to the statistical study of RHESSI spectra of non-flaring active regions by \citet{mctiernan2009}, the RHESSI-derived temperature for the active region reported here (7.1 to 7.4 MK) is well within the range that is reported by \citet{mctiernan2009} (6-10 MK), while the emission measure is at the lower end of the distribution. This is not surprising, as we were selecting a non-flaring active region near the detection limit of RHESSI. \citet{mctiernan2009} showed that the associated GOES level has generally a lower temperature between 3 and 6 MK with an emission measure that is generally larger by a factor of 50 to 100.  This difference of the GOES and RHESSI emission measures could be explained if the DEM has generally a shape as found in this work and the GOES flux would mainly originate from the cooler temperature plasma that has a much higher EM, while RHESSI sees the hotter tail of the distribution at lower EM. We can test this interpretation by estimating the GOES fluxes from the inferred DEMs shown in Figure 4. During orbit 1, the estimated GOES level from the derived DEM is around B1.4, with a standard deviation of B0.26 and extreme values of B1.0 and B2.1 for the different Monte Carlo runs. Assuming the flux from the active region dominates the GOES flux, these values are consistent with the observed value of B1.2. The isothermal GOES temperature derived from these  fluxes are 3.3 MK at an emission measure of 2.3$\times$10$^{48}$ cm$^{-3}$, similar to the main peak found the DEM shown in Figure 5. Hence, the difference between GOES and RHESSI quiescent active region observations could indeed by due to a DEM that generally has a peak around 2 to 3 MK and an hot tail. 

\subsection{NuSTAR}
For a quiescent active region investigation, NuSTAR and RHESSI have slightly different diagnostic capabilities. NuSTAR observes to lower photon energies than RHESSI with a peak in the count spectrum typically around $\sim$2 keV, while RHESSI has a peak in the count spectrum around $\sim$6 keV with a steeply decreasing sensitivity towards the detection limit at 3 keV. Hence, for a decreasing DEM in temperature, NuSTAR generally has a larger contribution from the cooler part of the distribution than RHESSI. Applying an isothermal fit to a decreasing DEM in temperature, NuSTAR will therefore give a lower temperature at a higher emission measure compared to RHESSI. \citet{hannah2016} reports on the detection of non-flaring active regions with NuSTAR up to $\sim$5 keV, without a signal at higher energies due to the limited throughput of NuSTAR. Isothermal fits to 5 quiescent active regions detected by NuSTAR over the 2.5 to 5 keV energy range reveal temperatures between 3.5 and 6 MK with emission measures in the range of several times 10$^{46}$ cm$^{-3}$. Compared to the DEMs shown in Figure 5 (right axis), the NuSTAR parameters are of the same order of magnitude as the discussed here indicating that the active regions investigated in the two studies are roughly of similar magnitude. If the DEM is indeed a steeply decaying function with temperature, the NuSTAR spectrum is actually determined by a range of temperatures with significant width, and not a single temperature. We estimate that for the DEM found for the active region in this paper, the 2.5 - 3.5 keV emission would actually be half from the distribution from plasma above and below 5 MK. Hence, NuSTAR non-flaring active region measurements would  benefit by adding measurements in EUV and SXR to independently estimate the contribution from colder plasmas. 

\subsection{FOXSI}

The DEMs presented here are similar to what was reported by \citet{ishikawa2017} using data from the second FOXSI sounding rocket flight. The main thermal components are around $10 ^{22}$~cm$^{-5}$ K$^{-1}$ for both studies, and the hot component DEM at 10~MK at with $\sim 10 ^{16}$~cm$^{-5}$ K$^{-1}$ is similar as well.  The FOXSI result includes a significant detection above 7~keV  making it a stronger result than the RHESSI results presented in this paper with the hottest temperatures detected being around 16~MK, compared to $\sim$8 MK for the study presented here. That both studies report similar temperature structures suggests that hot components are common in quiescent active regions.

\section{Summary}

Using RHESSI hard X-ray imaging spectroscopy observations during 4 different non-flaring intervals on July 13, 2010 separated by 18 hours, we were able to detect and measure the hot thermal tail in a quiescent active region in the range from 3.5 to 7.5 keV. RHESSI images reveal that the hot emission appears to be distributed over the entire active region. However, RHESSI imaging does not give us information about a filling factor. The detected emission could be volume-filling, as well as a sum of many compact nanoflares that are spatially separated by more than their individual extent.  

In combination with EUV and SXR narrow filter observations, we reconstructed the DEMs of a quiescent active region that show a peak between 2 and 3 MK and a steeply declining tail with temperature, similar as previously reported \citep{warren2012,delzanna2014,parenti2017}. The reconstructed DEMs strongly suggest that the RHESSI counts are produced by a range of temperatures ($\sim$3 to $\sim$8 MK). The standard approach of fitting a single temperature model to HXR spectra therefore introduces a bias, especially when energies below $\sim$6 keV are included in the fit range. As an example, we mention here that the isothermal fit gives temperatures between 7.1 and 7.4 MK when derived over the energy range from 3.5 to 7.5 keV, while the reconstructed DEM suggests that about 40\% of the 3.5-4.5 keV flux is produced by plasma below 5 MK. Hence, including EUV and SXR observations makes it clear that a single temperature approach is inadequate, despite that the isothermal fit to the HXR data alone represents the (few) data points well. 

To properly describe the DEM of quiescent active regions it is essential to combine EUV, SXR, and HXR observations. No wavelength range alone can cover the broad distribution of temperatures. The main thermal component dominates the observed flux in all the AIA EUV channels and in the available XRT filters. To measure the high temperature end, it is essential to have hard X-ray observations. For the active region discussed here, RHESSI only provides a detection below $\sim$7.5 keV, limiting the temperature diagnostics to below 8 MK. For a detection of the high temperature tail above 10 MK, hard X-ray observations above the 7.5 keV range are essential. The FOXSI Small Explorer mission concept has been designed to provide such measurements. 

While broadband filter observations such as XRT have greatly enhanced our understanding of hot plasmas in the solar corona, the next generation SXR telescopes should be imaging spectrometers. Such a telescope will provide spectra in SXR range that will allow us to directly fit continuum and line emissions for each pixel individually. In combination with EUV and HXR observations, SXR imaging spectroscopy will greatly enhance our knowledge of the DEM of active regions. Recently, the FOXSI-3 sounding rocket flight successfully performed SXR photon-counting imaging spectroscopy observations of the solar corona demonstrating the scientific potential and technical feasibility to build a satellite mission Physics of Energetic and Non-thermal plasmas in the X region (PhoENiX) with a similar SXR imaging spectroscopy instrument \citep{narukage2017}.

The existence of a hot tail in the DEM of quiescent active regions is an essential predication by nanoflare-heating models, but the details of the amount and distribution of the hot tail can vary depending on the model and parameters used. We encourage using our results for comparison of the results of numerical simulations of different nanoflare models. An obvious next step from the observational side is to perform a statistical study of quiescent active regions with already existing observations from RHESSI, AIA, and Hionode. The addition of EIS spectral data such as presented in \citet{warren2012} would significantly better constrain the main thermal component.

\acknowledgments
We thank the referee for the helpful comments, and we would like to thank Alicia Chavier for her help with the manuscript. This work was supported through KAKENHI grant 17H04832 from the Japan Society for the Promotion of Science. Additional support was provided through NASA contract NAS 5-98033 for RHESSI. 
A part of this study was carried out by using the computational resource of the Center for Integrated Data Science, Institute for Space-Earth Environmental Research, Nagoya University through the joint research program.
Hinode is a Japanese
mission developed and launched by ISAS/JAXA, with NAOJ
as domestic partner and NASA and STFC (UK) as international
partners. It is operated by these agencies in co-operation
with ESA and NSC (Norway).

\end{document}